\documentclass[11pt]{article}
\usepackage{amssymb}
\usepackage{epsfig}

\newcommand{\BE}{\begin{equation}}
\newcommand{\EE}{\end{equation}}
\begin{document}
\begin{titlepage}

\vspace*{1mm}
\begin{center}

\LARGE
   {\LARGE{\bf Basic randomness of nature and \\
   ether-drift experiments }}

\vspace*{14mm} {\Large  M. Consoli, A. Pluchino and A. Rapisarda}
\vspace*{10mm}\\
{\large
Istituto Nazionale di Fisica Nucleare, Sezione di Catania \\
Dipartimento di Fisica e Astronomia dell' Universit\`a di Catania \\
Via Santa Sofia 64, 95123 Catania, Italy \\ }
\end{center}
\begin{center}
{\bf Abstract}
\end{center}
We re-consider the idea that quantum fluctuations might reflect the
existence of an `objective randomness', i.e. a basic property of the
vacuum state which is independent of any experimental accuracy of
the observations or limited knowledge of initial conditions. Besides
being responsible for the observed quantum behaviour, this might
introduce a weak, residual form of `noise' which is intrinsic to
natural phenomena and could be important for the emergence of
complexity at higher physical levels. By adopting Stochastic Electro
Dynamics as a heuristic model, we are driven to a picture of the
vacuum as a form of highly turbulent ether, which is deep-rooted
into the basic foundational aspects of both quantum physics and
relativity, and to search for experimental tests of this scenario.
An analysis of the most precise ether-drift experiments, operating
both at room temperature and in the cryogenic regime, shows that, at
present, there is some ambiguity in the interpretation of the data.
In fact the average amplitude of the signal has precisely the
magnitude expected, in a `Lorentzian' form of relativity, from an
underlying stochastic ether and, as such, might not be a spurious
instrumental effect. This puzzle, however, should be solved in a
next future with the use of new cryogenically cooled optical
resonators whose stability should improve by about two orders of
magnitude. In these new experimental conditions, the persistence of
the present amplitude would represent a clean evidence for the type
of random vacuum we are envisaging.

\end{titlepage}

\section{Introduction}

The authors of Ref.\cite{random} have emphasized the possible
existence of an `objective randomness' as a basic property which is
independent of any experimental accuracy of the observations or
limited knowledge of initial conditions. In their opinion, this idea
is so important that quantum mechanics should be generalized or,
what is probably a more accurate perspective, should be recovered
within a new physical principle where randomness is taken as a
genuine property of nature. Actually, besides being responsible for
the observed quantum behaviour, this basic property might introduce
a residual form of {\it noise} that perturbs the system of interest
in a weak but unpredictable way.

If this were true, there might be important consequences. In fact,
it has becoming more and more evident that many classical and
quantum systems can increase their efficiency thanks to the presence
of noise. For example, it has been shown that noise-assisted
enhancement effects are crucial for both classical and quantum
communication channels. In this context, noise is supposed to play a
fundamental role in generating the quantum coherence that seems to
be involved in biological processes, such as pigment-protein complex
for photosynthesis in sulphur bacteria \cite{caruso}. But there are
other examples in which efficiency of classical systems is
reinforced by random noise, as for instance protein crystallization
\cite{frenkel}, noise enhanced stability \cite{spagnolo} or
stochastic resonance \cite{gamma1,gamma2}.

On this  basis, one is tempted to assume that the inclusion of an
objective noise, that reflects the effects of the environment and is
intrinsic to natural phenomena, might induce a new framework where
long-range correlations, complexity and also life,  emerge as a
natural consequence of underlying dynamical processes. In this
context, it is worthwhile to quote a new and general approach in
statistical mechanics, called {\it superstatistics}
\cite{beck-cohen1}, which deals with spatio-temporally fluctuating
intensive quantities in long-term stationary states of
non-equilibrium systems. Within this approach, a changing, noisy
environment, as when acting on a moving Brownian particle
\cite{beck-cohen2}, creates dynamical correlations which lead to a statistical
description where "fat-tailed" Probability Density Functions,
that characterize many complex systems, spontaneously emerge from a
superposition of local Gaussian distributions. For small amplitudes
of the fluctuations, such a behavior becomes universal and the
first-order corrections to the ordinary Boltzmann factor correspond
to those predicted by the so-called $q$-statistics introduced by
Tsallis in 1988 \cite{tsallis1,tsallis2}. These last considerations
reinforce the idea that a basic noise, at some elementary level,
could be crucial for the emergence of complexity at higher physical
levels.

Now, looking for an ultimate dynamical explanation, one could argue
as follows. If the required form of noise cannot be predicted or
controlled, it should be viewed as fundamentally simple. For the
same reason also the appropriate model environment, in spite of its
infinite number of degrees of freedom, may be considered as
basically simple. Therefore, in this paper, we shall concentrate on
the simplest possible state of any physical theory, the `vacuum',
and consider the following two questions that were left open in
Ref.\cite{random}:

1)  do the basic foundational aspects of quantum physics and
relativity point to some kind of random vacuum state?

2)  are there experimental signatures of this vacuum state that
might represent a fundamental form of  noise?

Exploring these two aspects represents a preliminary step in order
to take seriously the general framework illustrated above.

\section{Stochastic
Electro Dynamics and the idea of a turbulent ether}

Let us start to discuss question 1). Concerning the idea of an
objective randomness, an important motivation, that originates
within the quantum theory itself, was mentioned in
Ref.\cite{random}. Namely, one could try to modify the standard
deterministic evolution of the quantum states with regard to the
quantum theory of measurement and, in particular, to explain why
macroscopic objects are not observed in a superposition of states.
To this end, a number of models (for a complete review see
Ref.\cite{bassi}) was proposed to dynamically reduce the coherence
between macroscopically distinct states. In this context, randomness
plays a fundamental role in the original idea of Ghirardi, Rimini
and Weber \cite{ghirardi} of `spontaneous' localizations of the
microscopic systems. These processes should be considered a
consequence of the stochastic nature of space-time and are extremely
rare. However, they become important for a very large number of
elementary constituents because, then, macroscopic objects cannot
exist in a superposition of states for more than an infinitesimal
fraction of time. From an experimental point of view, the permanent
excitation of a body produced by the spontaneous localization
processes of its constituents represents, in a sense, a fundamental
form of noise that however is estimated to be too small to be
detected with present technology.

Alternatively, one could start from classical physics and assume
that the probabilistic aspects of quantum physics reflect the active
role of the vacuum whose stochastic nature modifies the classical
behaviour of the microsystems and provides the fundamental
background for the observed quantum fluctuations. This general idea,
shared by a large number of authors over the years (see e.g. the
long list reported in Ref.\cite{fritsche}), has produced various
formulations that sometimes differ non-trivially from each other and
could be denoted as `stochastic' or `hidden-variable' models of
quantum mechanics. In spite of the fact that, in some cases, the
predictions of these models can be made to agree with those of the
quantum theory, none of them can be considered a true `derivation'
of quantum mechanics from classical physics. However, this does not
mean that these models are useless. In fact, they might allow asking
questions that otherwise are not permissible in the standard quantum
theory (e.g. the origin of atomic stability, quantization,
indeterminacy,...).

Now, among such possible models, an interesting scenario is that of
Stochastic Electro Dynamics (SED)
\cite{boyer1,marshall1,marshall2,boyer2,delapena}. It provides a
definite classical framework that has genuine elements of randomness
and, by its very nature, points to the basic foundational aspects of
{\it both} quantum theory and relativity. As such, it will be
tentatively adopted in the following. By {\it tentatively} we mean
that SED is certainly not the only possible choice to discuss the
general idea of a random vacuum. Also, in agreement with the point
of view expressed by other authors \cite{rueda}, we do {\it not}
claim SED to be a complete, consistent theory. For instance, the
problems posed by a suitable generalization that might include the
existence of weak and strong interactions induce to give SED a
limited heuristic significance \footnote{Although limited, the
heuristic value of SED in our context reflects the fact that weak
and strong interactions were unknown at the beginning of 20th
century when both relativity and quantum physics were introduced. In
addition, the fact that the fine structure constant $\alpha=
{{e^2}\over{\hbar c}}\sim 1/137$ is a pure number means that
Planck's constant could also be expressed in terms of pure
electromagnetic constants \cite{nelsonbook0,davidson}.}.

However, SED provides an alternative derivation  of many physical
results such as the blackbody radiation spectrum, the fluctuations
in thermal radiation, the third law  of thermodynamics, rotator and
oscillator specific heats, the Van der Waals forces between
macroscopic objects and between polarizable particles (see
\cite{boyer2} and references quoted therein).

At the same time, the central premise of SED, which is relevant for
our purpose, is that the quantum behaviour of particles can {\it
also} be understood as the result of their classical interactions
with a vacuum, random radiation field. This field, considered in a
stationary state, is assumed to permeate all space and its  action
on the particles impresses upon them a stochastic motion with an
intensity characterized by Planck's constant. In this way, one can
get insight into basic aspects of the quantum theory such as the
wave-like properties of matter, indeterminacy, quantization,... For
instance, in this picture, atomic stability would originate from
reaching that `quantum regime' \cite{puthoffH,backto} which
corresponds to a dynamic equilibrium between the radiation emitted
in the orbital motions and the energy absorbed in the highly
irregular motions impressed by the vacuum stochastic field.

The theoretical framework of SED corresponds to the classical
Lorentz-Dirac theory \cite{dirac38}. Thus, for instance, an electron
in the field of a nucleus (in the non relativistic limit) is
described by the equation of motion \cite{cole} \BE m {{d^2 {\bf r}
}\over{ dt^2}}= -{{Ze^2 {\bf r}}\over{r^3}} - e~ [{\bf E} +
{{1}\over{c}}{{d {\bf r} }\over{ dt}}~ \times ~{\bf B} ] + {\bf
F}_{\rm reaction} \EE where the back reaction of the `ether'
\footnote{Davidson \cite{davidson0} argues that the peculiar aspects
of quantum mechanics, as embodied in the Schr\"odinger equation,
could be traced back to a statistical description of the radiative
reactive force of classical electromagnetism.} can be approximated
as (see e.g. \cite{panofski,rohrlich}) \BE {\bf F}_{\rm
reaction}\sim {{2}\over{3}} {{e^2}\over{c^2}} {{d^3 {\bf r} }\over{
dt^3}} \EE and where ${\bf E}$ and ${\bf B}$ represent the electric
and magnetic fields acting on the electron and include the
`zero-point' contributions \BE {\bf E}_{\rm ZP}({\bf r},t)=
{{1}\over{(L_xL_yL_z)^{1/2}}}\sum_{n_x,n_y,n_z=-\infty}^{+\infty}~\sum_{\lambda=1,2}\hat{\bf
\epsilon}_{{\bf k_n},\lambda}~ { f}_{ {\bf k_n}, \lambda}({\bf
r},t)\EE \BE {\bf B}_{\rm ZP}({\bf r},t)=
{{1}\over{(L_xL_yL_z)^{1/2}}}\sum_{n_x,n_y,n_z=-\infty}^{+\infty}~\sum_{\lambda=1,2}({\bf
k_n}~\times~\hat{\bf \epsilon}_{{\bf k_n},\lambda}) ~{ f}_{ {\bf
k_n}, \lambda}({\bf r},t) \EE with \BE f_{ {\bf k_n},\lambda}({\bf
r},t)= { a}_{ {\bf k_n},\lambda}\cos({\bf k_n}\cdot{\bf
r}-\omega_{\bf n}t) +{ b}_{{\bf k_n},\lambda}\sin( {\bf
k_n}\cdot{\bf r}-\omega_{\bf n}t) \EE Here $L_x$, $L_y$, $L_z$
denote the linear dimensions of the system, $n_x$, $n_y$, $n_z$ are
relative integers, ${\bf k_n}\equiv  2\pi ({{n_x}\over{L_x}},
{{n_y}\over{L_y}}, {{n_z}\over{L_z}})$, $\omega_{\bf n}=c|{\bf
k_n}|$  and the polarization vectors satisfy the conditions ${\bf
k_n}\cdot \hat{\bf \epsilon}_{{\bf k_n},\lambda}=0$ and $\hat{\bf
\epsilon}_{{\bf k_n},\lambda}\cdot \hat{\bf \epsilon}_{{\bf
k_n},\lambda'}=0$ for $\lambda \neq \lambda'$. Finally, the
coefficients ${ a}_{{\bf k_n},\lambda}$ and ${ b}_{{\bf
k_n},\lambda}$ in the plane wave expansion represent independent
random variables of the type that could be simulated by a random
number generator routine with zero mean and second moment
distributions \BE \langle{a}^2_{{\bf k_n},\lambda}\rangle=\langle{
b}^2_{{\bf k_n},\lambda}\rangle=2\pi \hbar\omega_{\bf n}\EE in order
to guarantee a Lorentz-invariant energy spectrum
$\rho_{ZP}(\omega)={{\hbar \omega^3}\over{(2\pi c^3)}}$. In this
sense, SED could be considered the same Lorentz classical electron
theory with new boundary conditions and it is remarkable that
numerical simulations \cite{cole} lead to electron trajectories that
nicely agree with the probability density of the Schr\"odinger wave
equation for the ground state of the hydrogen atom.

Even though many aspects have still to be understood (e.g. the
existence of metastable states corresponding to the higher energy
levels) we can draw the following conclusion. Usually, Lorentz
theory is only considered in connection with the origin of
relativity. However, within SED, it also provides interesting
insights on the quantum phenomena. To this end, one has simply to
replace the vanishing field used to characterize the lowest energy
state with a random zero-point field. But, then, this means that we
should change the picture of the Lorentz ether. Apparently, it
should no longer be thought as a stagnant fluid (for an observer at
rest) or as a fluid in laminar motion (for an observer in uniform
motion). Rather the ether should resemble a fluid in a chaotic
state, a fluid in a state of {\it turbulent} motion.

This same idea of an underlying turbulent ether is also supported by
other arguments. For instance, Maxwell equations can be derived
formally as hydrodynamic fluctuations of an incompressible turbulent
fluid \cite{troshkin,marmanis,puthoff2,tsankov}. As in the original
model proposed by Kelvin \cite{whittaker}, the energy which is
locally stored into the vortical motion becomes a source of
elasticity and the fluid resembles an elastic medium that can
support the propagation of transverse waves. In this derivation, one
starts from the Navier-Stokes or Euler equations for fluid dynamics
and splits the full velocity field  into  average $\langle
{v}_i\rangle $ and fluctuation component ${v'}_i$ . The existence of
transverse waves depends crucially on the Reynolds stress tensor
$\tau_{ik}=\langle v'_i v'_k\rangle$ which vanishes for a pure
laminar regime and, in this context, could be considered the analog
of the stress tensor of elastic media. Notice that, here, one starts
from a non-relativistic framework while Maxwell equations and their
Lorentz invariance `emerge' from the dynamics of the underlying
turbulent fluid.

A similar picture is also suggested by the formal equivalence
\cite{marcinkowski,kosevicbook} (velocity potential vs.
displacement, velocity vs. distortion, vorticity vs. density  of
dislocations,...) that can be established between various systems of
screw dislocations in an elastic solid and corresponding vortex
fields in a liquid. In this way, the phenomenon of turbulence can
provide a conceptual transition from fluid dynamics to a different
realm of physics, that of elasticity, where the wave speed, that by
itself is simply a quantity that remains invariant under changes of
the {\it average} velocity of the fluid, acquires also the meaning
of a limiting speed. This is due to the behaviour of the elastic
energy of moving dislocations (taken as models for the ordinary
ponderable matter) that increases proportionally to
$(1-v^2/c^2)^{-1/2}$, see e.g. \cite{frank}$-$\cite{christov}. This
type of correspondence, between turbulent fluids and elastic media,
leads to that intuitive visualization of the relativistic effects
which is characteristic of a Lorentzian approach.

Finally, this idea of a turbulent ether is also natural to get in
touch with other `stochastic' or `hidden-variable' models of quantum
mechanics. For instance, let us consider Nelson's mechanics. His
starting point is that ``particles in empty space, or let us say the
ether, are subject to Brownian motion" \cite{nelson}. However, the
fundamental nature of the phenomenon requires an ether with
vanishingly small friction \footnote{Actually Nelson assumed an
exactly zero-friction ether where the diffusion coefficient in the
Brownian motion for a particle with mass $m$ is $\nu=\hbar/(2m)$.
The case of an infinitesimal friction was studied by Kaloyerou and
Vigier \cite{friction}. This effectively amounts to the replacement
$\nu \to \nu e^{-\beta}$ and, by comparing with precision
experiments (e.g. the Lamb shift), one finds $|\beta| \lesssim
10^{-13}$.}``for then we could distinguish absolute rest from
uniform motion" \cite{nelson}. In this sense, apparently, the ether
should behave as a perfect vacuum. But, then, why there should be a
Brownian motion? A solution of this apparent contradiction can be
obtained if we use Onsager's original result \cite{onsager} on
turbulent fluids: in the zero-viscosity limit, i.e. infinite
Reynolds number, the fluid velocity field does {\it not} remain a
differentiable function. This irregular behaviour of the underlying
ether gives a physical argument to expect that the resulting
particle `` Brownian motion will not be smooth" \cite{nelson} and
thus to consider the particular form of kinematics which is at the
basis of Nelson's stochastic derivation of the Schr\"odinger
equation.

Analogously, in spite of some differences \cite{physrep}, this idea
of a fluid with very irregular and effectively random fluctuations
had also been advocated by Bohm and Vigier \cite{bohmvigier}.

On the other hand, it is also true that relativistic effects can be
described without ever mentioning the idea of an ether and the
methods of stochastic quantization \cite{parisi,huffel} can be
introduced as a pure theoretical construct \footnote{Another notable
exception is represented by Calogero's semi-quantitative derivation
\cite{calogero}. This is based on the chaotic structure of many-body
classical systems and the long-range nature of the gravitational
$1/r$ potential. As a consequence of these facts, in addition to the
standard gravitational effects, every particle should experience
locally a stochastic acceleration field (due to the rest of the
Universe) which, remarkably, appears to have the right order of
magnitude to explain the value of $\hbar$. To fill the gap with the
idea of an underlying stochastic medium, the interesting connections
between Brownian motion and potential theory \cite{knapp} could be
important.}. In this sense, one could adopt Nelson's words ("I
simply do not know whether the things I have been talking about are
physics or formalism" \cite{nelsonbook2}) to conclude that the idea
of a turbulent ether, although plausible, cannot be demonstrated on
the basis of the previous arguments.

Instead, to find definite support, one could look for some
unexpected experimental signature, thus coming to our question 2).
But what kind of experiment could ever detect a zero-viscosity
fluid? Up to now, the implicit assumption made by all authors is
that a `subquantal' ether (if any) is so elusive that its existence
can only be deduced {\it indirectly}, i.e. through the deviation of
the microsystems from the classical behaviour. However, what about
those highly sensitive `ether-drift' experiments that, since the
original Michelson-Morley experiment, have deeply influenced our
vision of relativity? Do they show any evidence for a non zero
effect? Eventually, could a tiny ether-drift represent the
manifestation of that fundamental form of noise we have envisaged?

At first sight, this possibility may seem in blatant contradiction
with Lorentz transformations. However, this is not necessarily true.
In fact, the speed of light in the vacuum, say $c_\gamma$, might not
coincide {\it exactly} with the basic parameter $c$ entering Lorentz
transformations, see e.g. Ref.\cite{uzan}. For instance, as stressed
in \cite{gerg}, this could happen in an `emergent-gravity' scenario
where, as in our case, one tends to consider the physical
vacuum as being not trivially `empty'. In this framework, the
space-time curvature observed in a gravitational field could
represent an effective phenomenon, analogously to a hydrodynamic
description of moving fluids on length scales that are much larger
than the size of the elementary constituents of the fluid
\cite{barcelo1,barcelo2,ultraweak}. Thus, although space-time is
exactly flat at the very fundamental level, one might be faced with
forms of curved `acoustic' metrics in which $c_\gamma \neq c$ thus opening
the possibility of a tiny but non-zero ether drift.

Here we want to emphasize that, in a scenario where one is taking
seriously a model of turbulent ether, there might be non trivial
modifications in the interpretation of the data. In fact, in the
traditional analysis of the ether-drift experiments, the
hypothetical, preferred reference frame associated with the ether
has always been assumed to occupy a definite, fixed location in
space. This induces to search for smooth time modulations of the
signal that might be synchronous with the Earth's rotation and its
orbital revolution. However, suppose that the ether were indeed
similar to a turbulent fluid. On the one hand, this poses the
theoretical problem of how to relate the macroscopic motions of the
Earth's laboratory (daily rotation, annual orbital revolution,...)
to the microscopic measurement of the speed of light inside the
optical cavities. On the other hand, from an experimental point of
view, it suggests sizeable random fluctuations of the signal that
could be erroneously interpreted as a mere instrumental effect.
Since the ultimate implications of our continuous flowing in such a
medium could be substantial, we believe that it is worth to
re-discuss these experiments in some detail by providing the reader
with the essential ingredients for their interpretation . After all,
other notable examples are known (e.g. the Cosmic Microwave
Background Radiation) where, at the beginning, an important physical
signal was interpreted as a mere instrumental effect.

In the following, we shall first review in Sect.3 the motivations to
re-propose a modern form of Lorentzian relativity, in connection
with the emergent-gravity scenario, and in Sect.4 the problem of
measuring the speed of light in vacuum optical cavities placed on
the Earth' surface. More technical aspects will be discussed in
Sects.5 and 6. These aspects are essential to fully appreciate the
puzzle posed by the present experimental situation: is the observed
signal a spurious instrumental effect or a non-trivial physical
manifestation of an underlying stochastic ether? Finally, Sect.7
will contain a summary and our conclusions with an outlook on the
planned experimental improvements.

\section{Lorentzian relativity and the emergent-gravity scenario}

There is a basic controversy about relativity that dates back to its
origin and concerns the interpretation of Lorentz transformations.
Do they originate from the {\it relative} motion of any pair of
observers $S'$ and $S''$, as in Einstein's special relativity, or
from the {\it individual} motion of each observer with respect to a
hypothetical preferred reference frame $\Sigma$ as in the
Lorentz-Poincar\'e formulation  ? As pointed out by several authors,
see e.g. \cite{bell,brown,brownbook,pla}, there is
no simple answer to this question. In fact, Lorentz transformations
have a group structure. Thus if $S'$ were individually related to
$\Sigma$ by a Lorentz transformation with dimensionless velocity
parameter $\beta'=v'/c$ and $S''$ were related to $\Sigma$ by a
Lorentz transformation with parameter $\beta''=v''/c$, the two
frames $S'$ and $S''$ would also be mutually connected by a Lorentz
transformation with relative velocity parameter \BE \beta_{\rm rel}=
{{\beta'-\beta''}\over{1-\beta'\beta''}}\equiv {{v_{\rm
rel}}\over{c}} \EE (we restrict for simplicity to one-dimensional
motions). This leads to a substantial quantitative equivalence of
the two formulations for most standard experimental tests where one
just compares the relative measurements of a pair of observers
\footnote{A clean and authoritative statement of this substantial
experimental equivalence could already be found in Ehrenfest's
inaugural lecture \cite{ehrenfest} held in Leyden on December 4th,
1912 "So, we see that the ether-less theory of Einstein demands
exactly the same here as the ether theory of Lorentz. It is, in
fact, because of this circumstance, that according to the Einstenian
theory an observer must observe the exact same contractions, changes
of rates, etc. in the measuring rods, clocks etc. moving with
respect to him as according to the Lorentzian theory. And let it be
said here right away in all generality. As a matter of principle,
there is no experimentum crucis between these two theories".}.

But now, what about ether-drift experiments ? In this context, the
basic issue concerns the value of $c_\gamma$, the speed of light in
the vacuum. Does it coincide {\it exactly} \cite{uzan} with the
basic parameter $c$ entering Lorentz transformations? Up to now, the apparent
failure of all attempts to measure the individual $\beta'$,
$\beta''$,... has been interpreted as an
experimental indication for $c_\gamma=c$ and this has provided,
probably, the main motivation for the wide preference given today to
special relativity.

However, if $c_\gamma= c$, also in the Lorentz-Poincar\'e
formulation relativistic effects conspire to make undetectable a
state of absolute motion in Michelson-Morley experiments. Therefore,
it is only the conceptual relevance of retaining a physical
substratum in the theory that may induce to re-discover the
potentially profound implications of the `Lorentzian' approach and
explore scenarios with tiny effects producing $c_\gamma\neq c$. To
this end, as anticipated, one could consider the emergent-gravity
scenario \cite{barcelo1,barcelo2} where the space-time curvature
observed in a gravitational field becomes an effective phenomenon,
analogously to a hydrodynamic description of moving fluids.

In this perspective, local distortions of the underlying ethereal
medium could produce local modifications of the basic space-time
units which are known, see e.g. \cite{feybook,dicke1}, to represent
an alternative way to generate an effective non-trivial curvature.
This point of view has been vividly represented  by K. Thorne in one
of his books \cite{thorne}: "Is space-time really curved ? Isn't
conceivable that space-time is actually flat, but clocks and rulers
with which we measure it, and which we regard as perfect, are
actually rubbery ? Might not even the most perfect of clocks slow
down or speed up and the most perfect of rulers shrink or expand, as
we move them from point to point and change their orientations ?
Would not such distortions of our clocks and rulers make a truly
flat space-time appear to be curved ? Yes".

By following this type of interpretation, one could first consider a
simplest two-parameter scheme \cite{ultraweak} in which there are
simultaneous re-scalings of i) any mass $m$ (and binding energy) and
of ii) the velocity of light in the vacuum as with a non-trivial
vacuum refractive index, i.e. \BE m\to
\hat{m}(x)~~~~~~~~~~~~~~~c_\gamma\to {{c}\over{{\cal{N}}(x)}} \EE In
this case, the physical units would also be rescaled \BE
\hat{t}(x)={{\hbar}\over{\hat{m}(x)c^2}}\equiv\lambda(x)t
~~~~~~~~~~~~~~~~~
\hat{l}(x)={{\hbar}\over{\hat{m}(x)c}}\equiv\lambda(x)l \EE
producing the effective metric structure ($A=c^2
{{\lambda^2}\over{{\cal{N}}^2}}$ and $B=\lambda^2$)\BE \label{first}
{g}_{\mu \nu}= {\rm diag}(A ,-B,-B,-B)\EE whose consistency with
experiments requires the weak-field identification with the
Newtonian potential \BE \label{second} {\cal{N}}\sim
1+2{{|U_N|}\over{c^2}}~~~~~~~~~~~~~~ \lambda\sim
1+{{|U_N|}\over{c^2}}\EE Then, more complicated metrics with
off-diagonal elements $g_{0i}\neq 0$ and $g_{ij}\neq 0$ can be
obtained by applying boosts and rotations to Eq.(\ref{first}) thus
basically reproducing the picture of the curvature effects in a
moving fluid. In this way, one is driven to consider the possibility
of a non-zero (but admittedly extremely small) light anisotropy that
could be measured in the present generation of precise ether-drift
experiments. This other part will be discussed in the following
section.

\section{The speed of light in the vacuum}

After having discussed why gravity might induce local modifications
of the basic space-time units, let us now consider the problem of
measuring the speed of light. On a general ground, to determine
speed as (distance moved)/(time taken), one must first choose some
standards of distance and time. Since different choices can give
different answers, we shall adopt in the following the point of view
of special relativity where the speed of light in the vacuum
$c_\gamma$, when measured in an inertial frame, coincides with the
basic parameter $c$ that enters Lorentz transformations. However,
inertial frames are just an idealization. Therefore the appropriate
realization is to assume {\it local} standards of distance and time
such that the identification $c_\gamma=c$ holds as an asymptotic
relation in the physical conditions which are as close as possible
to an inertial frame, i.e. {\it in a freely falling frame} (at least
by restricting to a space-time region small enough that tidal
effects of the external gravitational potential $U_{\rm ext}(x)$ can
be ignored). This is essential to obtain an operative definition of
the otherwise unknown parameter $c$. At the same time, the
consistency of this scheme can be checked by comparing with
experiments.

In fact, with these premises, light propagation for an observer $S'$
sitting on the Earth's surface can be described with increasing
degrees of approximations \cite{gerg}:

\begin{figure}
\begin{center}
\epsfig{figure=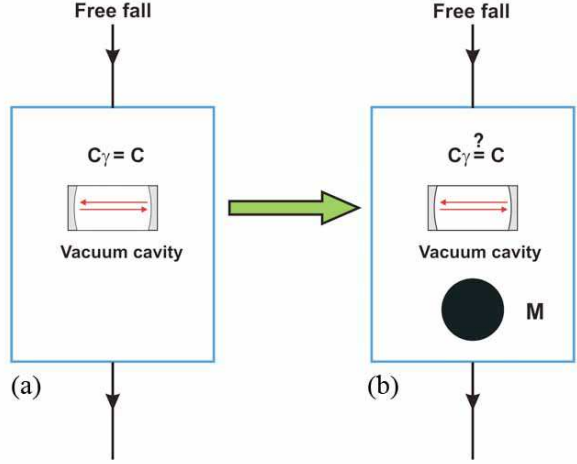,width=8.5truecm,angle=0}
\end{center}
\caption{ {\it A pictorial representation of the effect of a heavy
mass $M$ carried on board of a freely-falling system, case (b). With
respect to case (a), in a flat-space picture of gravity, the mass
$M$ modifies the effective, local space-time structure by re-scaling
the physical units ($dx$, $dy$, $dz$, $dt$) $\to$ ($d\hat{x}$,
$d\hat{y}$, $d\hat{z}$, $d\hat{t}$) and introducing a non-trivial
refractive index ${\cal N}\neq 1$ so that now $c_\gamma \neq c$}. }
\label{Fig.1}
\end{figure}

~~~i) $S'$ is considered a freely falling frame. This amounts to assume
$c_\gamma=c$ so that,
given two events which, in terms of the local space-time units of
$S'$, differ by $(dx, dy, dz, dt)$, light propagation is described
by the condition (ff='free-fall') \BE\label{zero1} (ds^2)_{\rm
ff}=c^2dt^2- (dx^2+dy^2+dz^2)=0~\EE ~~~ii) Now, is really the Earth a
freely-falling frame ? To a closer look, in fact, an observer  $S'$
placed on the Earth's surface can only be considered a
freely-falling frame up to the presence of the Earth's gravitational
field. Its inclusion leads to tiny deviations from the standard
Eq.(\ref{zero1}). These can be estimated by considering $S'$ as a
freely-falling frame (in the same external gravitational field
described by $U_{\rm ext}(x)$) that however is also carrying on
board a heavy object of mass $M$ (the Earth's mass itself) that
affects the effective local space-time structure (see Fig.1). To derive the
required correction, let us again denote by ($dx$, $dy$, $dz$, $dt$)
the local space-time units of the freely-falling observer $S'$ in
the limit $M=0$ and by $\delta U$ the extra Newtonian potential
produced by the heavy mass $M$ at the experimental set up where one
wants to describe light propagation. From Eqs.(\ref{first}) and
(\ref{second}), in an emergent-gravity scenario, light propagation
for the $S'$ observer can then be described by the condition
\cite{gerg}\BE\label{iso}(ds^2)_{\rm \delta U} ={{c^2d\hat{t}
^2}\over{{\cal N}^2 }}- (d\hat{x}^2+d\hat{y}^2+d\hat{z}^2)=0~\EE
where, to first order in $\delta U$, the space-time units
($d\hat{x}$, $d\hat{y}$, $d\hat{z}$, $d\hat{t}$) are related to the
corresponding ones ($dx$, $dy$, $dz$, $dt$) for $\delta U=0$ through
an overall re-scaling factor \BE \label{lambda} \lambda= 1+{{|\delta
U|}\over{c^2}} \EE and we have also introduced the vacuum refractive index
\BE\label{refractive1}{\cal N}= 1+2{{|\delta
U|}\over{c^2}} \EE Therefore, to this order, light is formally
described as in General Relativity where one finds the weak-field,
isotropic form of the metric \BE\label{gr} (ds^2)_{\rm
GR}=c^2dT^2(1-2{{|U_{\rm N}|}\over{c^2}})-
(dX^2+dY^2+dZ^2)(1+2{{|U_{\rm N}|}\over{c^2}})\equiv c^2 d\tau^2 -
dl^2\EE In Eq.(\ref{gr}) $U_N$ denotes the Newtonian potential and
($dT$, $dX$, $dY$, $dZ$) arbitrary coordinates defined for $U_{\rm
N}=0$. Finally, $d\tau$ and $dl$ denote the elements of proper time
and proper length in terms of which, in General Relativity, one
would again deduce from $ds^2=0$ the same universal value
$c={{dl}\over{d\tau}}$. This is the basic difference with
Eqs.(\ref{iso})-(\ref{refractive1}) where the physical unit of
length is $\sqrt {d\hat{x}^2+d\hat{y}^2+d\hat{z}^2}$, the physical
unit of time is $d\hat{t}$ and  instead a non-trivial refractive
index ${\cal N}$ is introduced. For an observer placed on the
Earth's surface, its value is \BE \label{refractive}{\cal N}- 1 \sim
{{2G_N M}\over{c^2R}} \sim 1.4\cdot 10^{-9}\EE  $M$ and $R$ being
respectively the Earth's mass and radius.

~~~iii) Differently from General Relativity, in a flat-space
interpretation with re-scaled units ($d\hat{x}$, $d\hat{y}$,
$d\hat{z}$, $d\hat{t}$) and ${\cal N}\neq 1$, the speed of light in
the vacuum $c_\gamma$ no longer coincides with the parameter $c$
entering Lorentz transformations. Therefore, as a general
consequence of Lorentz transformations, an isotropic propagation as
in Eq.(\ref{iso}) can only be valid for a special state of motion of
the Earth's laboratory. This provides the operative definition of a
preferred reference frame $\Sigma$ while for a non-zero relative
velocity ${\bf V}$  one expects off diagonal elements $g_{0i}\neq 0$
in the effective metric and a tiny light anisotropy. As shown in
Ref.\cite{gerg}, to first order in both $({\cal N}- 1)$ and $V/c$
one finds \BE g_{0i}\sim 2({\cal N}- 1){{V_i}\over{c}} \EE These off
diagonal elements can be imagined as being due to a directional
polarization of the vacuum induced by the now moving Earth's
gravitational field and express the general property \cite{volkov}
that any metric, locally, can always be brought into diagonal form
by suitable rotations and boosts. In this way, by introducing
$\beta=V/c$, $\kappa=( {\cal N}^2 -1)$ and the angle $\theta$
between ${\bf V}$ and the direction of light propagation, one finds,
to ${\cal O}(\kappa)$ and ${\cal O}(\beta^2)$, the one-way velocity
\cite{gerg}
\BE \label{oneway}
       c_\gamma(\theta)= {{c} \over{{\cal N}}}~\left[
       1- \kappa \beta \cos\theta -
       {{\kappa}\over{2}} \beta^2(1+\cos^2\theta)\right]
\EE
and a two-way velocity of light
\begin{eqnarray}
\label{twoway}
       \bar{c}_\gamma(\theta)&=&
       {{ 2  c_\gamma(\theta) c_\gamma(\pi + \theta) }\over{
       c_\gamma(\theta) + c_\gamma(\pi + \theta) }} \nonumber \\
       &\sim& {{c} \over{{\cal N}}}~\left[1-\beta^2\left(\kappa -
       {{\kappa}\over{2}} \sin^2\theta\right) \right]
\end{eqnarray}
This allows to define the RMS \cite{robertson,mansouri} anisotropy
parameter ${\cal B }$ through the relation \BE \label{rms}
    {{\Delta \bar{c}_\theta } \over{c}}=
    {{\bar{c}_\gamma(\pi/2 +\theta)- \bar{c}_\gamma (\theta)} \over
       {\langle \bar{c}_\gamma \rangle }} \sim{\cal B }
       {{V^2 }\over{c^2}} \cos(2\theta) \EE
with
\BE \label{rmsb}
       |{\cal B }|\sim {{\kappa}\over{2}}\sim {\cal N}-1
\EE
From the previous analysis, by replacing the value of the refractive
index Eq.(\ref{refractive}) and adopting, as a rough order of
magnitude, the typical value of most cosmic motions $V\sim$ 300
km/s \footnote{For instance, from the motion of the Solar System within the
Galaxy, or with respect to the centroid of the Local Group or with
respect to the CMBR, one gets respectively $V\sim$ 240, 320, 370 km/s.}
, one expects a tiny fractional anisotropy \BE \label{averani}
        {{\langle\Delta \bar{c}_\theta \rangle} \over{c}} \sim
       |{\cal B }|{{V^2 }\over{c^2}} ={\cal O}(10^{-15}) \EE
that could finally be detected in the present, precise ether-drift
experiments. These experiments will be discussed in the
following section.

\begin{figure}
\begin{center}
\epsfig{figure=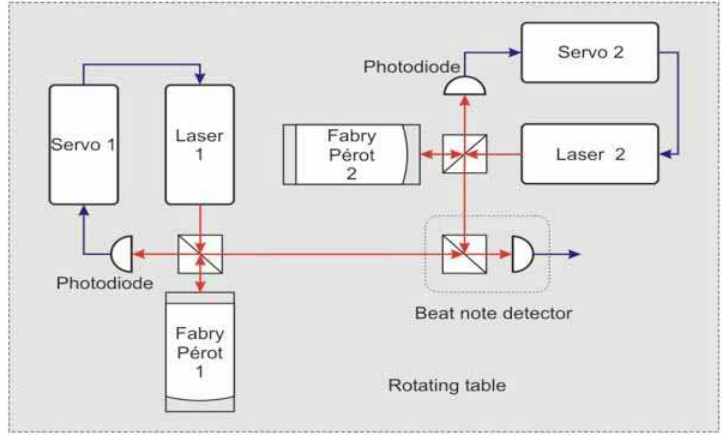,width=10truecm,angle=0}
\end{center}
\caption{ {\it The scheme of a modern ether-drift experiment. The
frequencies $\nu_1$ and $\nu_2$ of the signals from the two
Fabry-Perot resonators are compared in the beat note detector that
provides the frequency shift $\Delta \nu=\nu_1 -\nu_2$. In this
picture, the apparatus is fully symmetric. On the other hand, in
Ref.\cite{peters} only one of the two resonators was rotating while
the other was kept fixed in the laboratory and oriented
north-south}. } \label{Fig.2}
\end{figure}

\section{Ether-drift experiments and stochastic ether}

In the present ether-drift experiments one measures the frequency shift,
i.e. the beat signal, $\Delta \nu$ of two cavity-stabilized lasers (see Fig.2)
whose definite non-zero value would provide a direct measure of an
anisotropy of the velocity of light \cite{applied}. In this
framework, the possible time modulation of the signal that might be
induced by the Earth's rotation (and its orbital revolution) has
always represented a crucial ingredient for the analysis of the
data. For instance, let us consider the relative frequency shift for
the experiment of Ref.\cite{peters}. Here the basic concept of light anisotropy
Eq.(\ref{rms}) as a second-harmonic effect leads to the
parametrization  \BE \label{basic2}
     {{\Delta \bar{c}_\theta(t) } \over{c}} = {{\Delta \nu (t)}\over{\nu_0}} =
      {S}(t)\sin 2\omega_{\rm rot}t +
      {C}(t)\cos 2\omega_{\rm rot}t
\EE
where $\nu_0$ indicates the reference frequency of the two resonators and
 $\omega_{\rm rot}$ is the rotation frequency of one resonator
with respect to the other which is kept fixed in the laboratory and
oriented north-south. If one assumes the picture of a {\it fixed} preferred frame
$\Sigma$ then, for short-time
observations of 1-2 days, the time dependence of a hypothetical
physical signal can only be due to (the variations of the projection
of the Earth's velocity ${\bf V}$ in the interferometer's plane
caused by) the Earth's rotation. In this case,
the two functions $S(t)$ and $C(t)$ admit the
simplest Fourier expansion \cite{peters} ($\tau=\omega_{\rm sid}t$ is the sidereal
time of the observation in degrees)  \BE
\label{amorse1}
      {S}(t) = S_0 +
      {S}_{s1}\sin\tau +{S}_{c1} \cos\tau
       + {S}_{s2}\sin(2\tau) +{S}_{c2} \cos(2\tau)
\EE \BE \label{amorse2}
      {C}(t) = {C}_0 +
      {C}_{s1}\sin\tau +{C}_{c1} \cos\tau
       + {C}_{s2}\sin(2\tau) +{C}_{c2} \cos(2\tau)
\EE
with {\it time-independent} $C_k$ and $S_k$ Fourier coefficients. Thus, by
accepting this theoretical framework, it becomes natural to average
the various $C_k$ and $S_k$ obtained from fits performed during a
1-2 day observation period. By further averaging over many
short-period experimental sessions, the data support the general
conclusion \cite{joint,newberlin,schillernew} that, although the
typical instantaneous $S(t)$ and $C(t)$ are ${\cal O}(10^{-15})$,
the global averages $(C_k)^{\rm avg}$ and $(S_k)^{\rm avg}$ for the
Fourier coefficients are much smaller, at the level ${\cal
O}(10^{-17})$, and, with them, the derived parameters entering the
phenomenological SME \cite{sme} and RMS \cite{robertson,mansouri}
models.

However, there might be different types of ether-drift where the
straightforward parameterizations Eqs.(\ref{amorse1}), (\ref{amorse2})
and the associated averaging procedures are {\it
not} allowed. Therefore we believe that, before assuming any definite
theoretical scenario, one should first ask: if light were really
propagating in a physical medium, an ether, and not in a trivial
empty vacuum, how should the motion of (or in) this medium be
described? Namely, could this relative motion exhibit variations
that are {\it not} only due to known effects as the Earth's rotation
and orbital revolution?

The point is that, by representing the physical vacuum as a fluid,
the standard assumption of smooth sinusoidal variations of the
signal, associated with the Earth's rotation (and its orbital
revolution), corresponds to assume the conditions of a pure laminar
flow associated with simple regular motions. Instead, by
adopting the model of an underlying turbulent medium there might
be other forms of time modulations. In this
alternative scenario, the same basic experimental data might admit a
different interpretation and a definite instantaneous signal $\Delta
\nu (t)\neq 0$ could become consistent with $(C_k)^{\rm avg} \sim
(S_k)^{\rm avg}\sim 0$.

To exploit the possible implications,
let us first recall the general aspects of any turbulent flow. This
is characterized by extremely irregular variations of the velocity,
with time at each point and between different points at the same
instant, due to the formation of eddies \cite{landau}. For this
reason, the velocity continually fluctuates about some mean value
and the amplitude of these variations is {\it not} small in
comparison with the mean velocity itself. The time dependence of a
typical turbulent velocity field can be expressed as \cite{landau}
\BE {\bf v}(x,y,z,t)=\sum_{p_1p_2..p_n} {\bf
a}_{p_1p_2..p_n}(x,y,z)\exp(-i\sum^{n}_{j=1}p_j\phi_j) \EE where the
quantities $\phi_j=\omega_j t+ \beta_j $ vary with time according to
fundamental frequencies $\omega_j$ and depend on some initial phases
$\beta_j$. As the Reynolds number ${\cal R}$ increases, the total
number $n$ of $\omega_j$ and $\beta_j$ increases thus suggesting a sequence
where laminar flow first becomes periodic, then quasi-periodic and finally
highly turbulent. In this limit, where ${\cal R}
\to \infty$, the required number of frequencies diverges
so that the theory of such
a turbulent flow must be a statistical theory.

Now, as anticipated in Sect.2, there are arguments to consider the
limit of an ether with vanishingly small viscosity where, indeed, the
relevant Reynolds numbers should become infinitely large in most
regimes. In this case, one is faced precisely with the limit of a
fully developed turbulence where the temporal analysis of the flow
requires an extremely large number of frequencies and the physical
vacuum behaves as a {\it stochastic} medium. Thus random
fluctuations of the signal, superposed on the smooth sinusoidal
behaviour associated with the Earth's rotation (and orbital
revolution), would produce deviations of the time dependent
functions $S(t)$ and $C(t)$ from the simple structure in
Eqs.(\ref{amorse1}) and (\ref{amorse2}) and an effective temporal
dependence of the fitted $C_k=C_k(t)$ and $S_k=S_k(t)$. In this
situation, due to the strong cancelations occurring in vectorial
quantities when dealing with stochastic signals, one could easily
get vanishing global inter-session averages \BE (C_k)^{\rm avg} \sim
(S_k)^{\rm avg} \sim 0\EE Nevertheless, as it happens with the
phenomena affected by random fluctuations, the average quadratic
amplitude of the signal could still be preserved. To this end, let
us re-write Eq.(\ref{basic2}) as \BE \label{basic3}
 {{\Delta \bar{c}_\theta(t) } \over{c}}={{\Delta \nu (t)}\over{\nu_0}} =
      A(t)\cos (2\omega_{\rm rot}t -2\theta_0(t))
\EE
where
\BE \label{interms}
C(t)=A(t)\cos2\theta_0(t)~~~~~~~~S(t)=A(t)\sin2\theta_0(t)\EE so that
\BE A(t)= \sqrt{S^2(t) +C^2(t)} \EE
Here $\theta_0(t)$ represents the instantaneous direction of a hypothetical
ether-drift effect in the x-y plane of the interferometer
(counted by convention from North through East so that North is
$\theta_0=0$ and East is $\theta_0=\pi/2$). By also introducing the
magnitude $v=v(t)$ of the
{\it projection} of the full ${\bf V}$, such that
\BE v_x(t)=v(t)\sin\theta_0(t)~~~~~~~~~~~~~~~v_y(t)=v(t)\cos\theta_0(t) \EE
  and adopting the same notations as in Eq.(\ref{averani}), we obtain the
  theoretical relations \cite{gerg}
\BE \label{amplitude1}
       A_{\rm th }(t)= {{1}\over{2}}|{\cal B }| {{v^2(t) }\over{c^2}}
\EE and \BE \label{amplitude10}
       C_{\rm th }(t)= {{1}\over{2}}|{\cal B }|~ {{v^2_y(t)- v^2_x(t)  }
       \over{c^2}}~~~~~~~~~~~~~~
       S_{\rm th }(t)= {{1}\over{2}}|{\cal B }| ~{{2v_x(t)v_y(t)  }\over{c^2}}
\EE In this way, in a stochastic ether, the positive-definite
amplitude $A(t)$ of the signal will have a definite non-zero average
value $\langle A\rangle$ and this can well coexist with $(C_k)^{\rm avg}
\sim (S_k)^{\rm avg}\sim 0$. Physical conclusions will then require
to first compare the measured value of $\langle A\rangle$ with the
short-term, stability limits of the individual optical resonators
and then with the theoretical expectation
(\ref{amplitude1}).

\section{Instrumental effects or fundamental noise?}

To provide evidence that indeed, in ether-drift experiments, we
might be faced with a form of fundamental noise from an underlying
stochastic ether, the present, most precise experiments
\cite{newberlin,crossed} were considered in Ref.\cite{gerg}. In the
experimental apparatus of Ref.\cite{crossed}, to minimize all
sources of systematic asymmetry, the two optical cavities were
obtained from the same monolithic block of ULE (Ultra Low Expansion
material). In these conditions, due to sophisticated electronics and
temperature controls, the short-term (about 40 seconds) stability
limits for the individual optical cavities are extremely high.
Namely, for the non-rotating set up, by taking into account various
systematic effects, one deduces stabilities $(\delta \nu)_1\sim
(\delta\nu)_2\sim  \pm 0.05$ Hz for the individual cavities 1 and 2
and thus about $\pm 2\cdot 10^{-16}$ in units of a laser frequency
$\nu_0=2.82\cdot 10^{14}$ Hz. This is of the same order of the {\it
average} frequency shift between the two resonators, say $(\Delta
\nu)^{\rm avg} \sim \pm 0.06$ Hz, when averaging the signal over a
very large number of temporal sequences (see their Fig.9b).

However, the magnitude of the {\it instantaneous} frequency shift
$\Delta\nu(t)$ is much larger, say $\pm 1$ Hz (see their Fig.9a),
and so far has been interpreted as a spurious instrumental effect.
To check this interpretation, we observe that, in the absence of any
genuine physical signal, the frequency shift between the two
resonators should exhibit the same typical instabilities $(\delta
\nu)_1$ and $(\delta\nu)_2$ of the individual resonators and thus,
for short-time observations, should be at the same level $\pm 2\cdot
10^{-16}$. Instead, for the same non-rotating set up, the minimum
noise in the frequency shift $\Delta\nu$ was found about 10 times
bigger, namely $1.9\cdot 10^{-15}$ (see Fig.8 of
Ref.\cite{crossed}). Also the trend of this form of noise in the
beat signal, as function of the averaging time, is different from
the corresponding one observed in the individual resonators thus
suggesting that the two types of noise might have different origin.

The authors tend to interpret this relatively large beat signal as
cavity thermal noise and refer to \cite{numata}. However, this
interpretation is not so obvious since the typical disturbances
$(\delta \nu)_1$ and $(\delta\nu)_2$ in the {\it individual} cavities  were
reduced to a considerably lower level.

For a quantitative estimate of the amplitude $A(t)$ of the signal we
can consider the more recent paper \cite{schillernew} of the same
authors. The physical second-harmonic part of the signal that
corresponds to an ether-drift effect, from their Eq.(1), can be
expressed as \footnote{For an easier comparison, we maintain the
notations of Ref.\cite{schillernew} where $B(t)$ is used to denote
the same amplitude $S(t)$ introduced before in Eq.(\ref{basic2}) and
the overall factor of 2 takes into account the differences with
respect to Eq.(\ref{basic2}) introduced by a fully symmetric
apparatus.} \BE \left({{\Delta \nu (t)}\over{\nu_0}}\right)^{\rm
physical} =
      2{B}(t)\sin 2\omega_{\rm rot}t +
      2{C}(t)\cos 2\omega_{\rm rot}t \equiv
      A^{\rm symm}(t) \cos(2\omega_{\rm rot}t
      -2 \theta_0(t))
\EE where \BE A^{\rm symm}(t) = 2\sqrt{B^2(t) +C^2(t)} \EE Now the
data of Ref.\cite{schillernew} confirm the above mentioned trend
with average values $\langle B\rangle$ and $\langle C\rangle$ which
are much smaller than their typical instantaneous values since one
finds (see their Fig.3)  \BE \langle B\rangle \sim \langle C\rangle
\sim {\cal O}(10^{-17}) \EE Therefore the quadratic average values
$\langle B^2\rangle$ and $\langle C^2 \rangle$ are essentially
determined by the variances  $\sigma_B\sim 7.5\cdot 10^{-16}$ and
$\sigma_C\sim 6.1\cdot 10^{-16}$ \cite{schillernew}. In this way, we
obtain the experimental value \BE \langle A^{\rm symm} _{\rm
exp}\rangle \sim {2}\sqrt{\sigma^2_B +\sigma^2_C} \sim 1.9 \cdot
10^{-15} \EE in good agreement with our theoretical expectation from
Eqs.(\ref{refractive}), (\ref{rmsb}) and (\ref{amplitude1}) for the
average Earth's velocity of most cosmic motions $\sqrt{\langle v^2
\rangle} \sim$ 300 {km/s}
 \BE \langle A^{\rm symm} _{\rm th}\rangle =
2\langle A _{\rm th}\rangle=|{\cal B }| {{\langle v^2\rangle }\over{c^2}}
 \sim 1.4 \cdot 10^{-15} {{\langle v^2 \rangle
}\over{(300 ~ \rm {km/s})^2}} \EE Similar conclusions can be
obtained from the other experiment of Ref.\cite{newberlin} where the
stability of the individual resonators is at the same level of a few
$10 ^{-16}$. Nevertheless, the measured $C(t)$ and $S(t)\equiv B(t)$
entering the beat signal span the whole range $ \pm 1.2\cdot
10^{-15}$ (see their Fig.4a) and are again interpreted in terms of a
thermal noise of the individual cavities. Thus, in the present two
most precise ether-drift experiments, the average amplitude of the
signal is about 4-5 times larger than the short-term stability of
the individual resonators and its measured value $ \langle A\rangle=
{\cal O}(10^{-15})$ is completely consistent with our theoretical
expectations.

Finally, as an additional check, a comparison with a previous
experiment \cite{schiller} operating in the cryogenic regime was
also performed in Ref.\cite{gerg}. Again, by restricting to the
variable part of the signal which is less affected by spurious
systematic effects (see Ref.\cite{gerg}), the average amplitude was
found $ {\cal O}(10^{-15})$. Thus this stable value of about
$10^{-15}$ found in all experiments is unlike to represent a
spurious instrumental artifact of the individual optical cavities of
the type discussed in Ref.\cite{numata}. In fact, the estimate of
Ref.\cite{numata} is based on the fluctuation-dissipation theorem,
and therefore there is no reason that both room temperature and
cryogenic experiments exhibit the same instrumental noise. This
argument confirms that, at present, there is a basic ambiguity in
the interpretation of the experimental data. The standard
interpretation in terms of spurious instrumental effects of the
individual optical cavities is by no means unique and the observed
signal could also represent a fundamental noise associated with the
underlying stochastic ether.

The puzzle, however, should be definitely solved in a next future.
In fact, the authors of Ref.\cite{newberlin} are starting to upgrade
their apparatus with cryogenically cooled sapphire optical  cavities
\cite{upgrading}. This should improve the short-term stability of
the individual resonators by about two orders of magnitude (say well
below the $10^{-17}$ level). In these new experimental conditions,
the persistence of an average amplitude $ \langle A\rangle= {\cal
O}(10^{-15})$ (i.e. about 100 times larger) would represent an
unambiguous evidence for the type of random vacuum we have been
considering.

\section{Summary and conclusions}

In this paper, by following the authors of Ref.\cite{random}, we
have re-considered the idea of an `objective randomness' in nature
as a basic property, independent of any experimental accuracy of the
observations or limited knowledge of initial conditions. The
interesting aspect is that, besides being responsible for the
observed quantum behaviour, this might introduce a weak, residual
form of noise which is intrinsic to natural phenomena and could be
important for the emergence of complexity at higher physical levels,
as suggested by both theoretical and phenomenological evidence.

By trying to implement this idea in a definite dynamical framework,
and adopting Stochastic Electro Dynamics as a heuristic model, we
have been driven to the idea of the vacuum as an underlying
turbulent ether which is deep-rooted into the basic foundational
aspects of {\it both} quantum physics and relativity. Thus,  by
searching for experimental tests of this scenario, we have
concentrated on the modern ether-drift experiments. Our analysis of
the most precise experiments (operating both at room temperature and
in the cryogenic regime) shows that, at present, there is some
ambiguity in the interpretation of the data. In fact, the average
amplitude of the beat signal between the two resonators has
precisely the magnitude expected, in a `Lorentzian' form of
relativity, from an underlying stochastic ether and, as such, might
not be a spurious instrumental effect but the manifestation of that
fundamental form of noise we have envisaged.

This puzzle, however, should be definitely solved in a next future
with the use of new cryogenically cooled optical cavities whose
individual stability should improve by about two orders of
magnitude. In these new experimental conditions, the persistence of
the present amplitude would represent an unambiguous evidence for
the type of random vacuum we have been considering. Namely, this
would turn out to be similar to a polarizable medium, responsible
for the apparent curvature effects seen in a gravitational field
and, at the same time, a stochastic medium, similar to a
zero-viscosity fluid in a turbulent state of motion, responsible for
the observed strong random fluctuations of the signal. All together,
the situation might resemble the discovery of the Cosmic Microwave
Background Radiation that, at the beginning, was also interpreted as
a mere instrumental effect.

Of course, a long series of steps will further be needed to explore
the ultimate implications of such new possible framework. For
instance, a confirmation of our prediction means that, in agreement
with the intuitive notion of an `ether wind', our continuous flowing
in the physical vacuum produces an infinitesimal energy-momentum
flux. Therefore, after its unambiguous detection `in vacuum', in the
completely controlled conditions of the announced, new generation of
cryogenic experiments, one might also consider different tests. For
instance, ether-drift experiments where cavities are filled by
different forms of matter \cite{muller2,epjc} that represent a
useful, complementary tool to study small deviations from exact
Lorentz invariance. The combined informations from this set of
observations of the characteristics of the signal (time modulations,
intensity, spatial coherence,...)  will be crucial to understand if
this tiny effect, by exposing all physical systems to an energy
flux, can induce the spontaneous generation of order which is
important for the emergence of complexity.
\\
\centerline{\bf Acknowledgments}

\noindent We thank L.Pappalardo for useful discussions and collaboration.

\end{document}